\documentclass[amsmath,amssymb,12pt,singlespacing,longbibliography,onecolumn,tightenlines]{revtex4-2}
\usepackage{epsfig}
\usepackage{graphicx}
\usepackage{bm}
\usepackage{mathrsfs}
\usepackage{xcolor}
\usepackage{verbatim}
\makeatletter
\def\mathcolor#1#{\@mathcolor{#1}}
\def\@mathcolor#1#2#3{%
  \protect\leavevmode
  \begingroup
    \color#1{#2}#3%
  \endgroup
}
\makeatother
\usepackage{comment}
\usepackage{cancel}
\usepackage{tensor}
\usepackage{physics}
\usepackage{siunitx}
\usepackage{soul}
\newcommand{\hq}{h_{\scriptscriptstyle  ?}}
\newcommand{\hbarq}{\hbar_{\scriptscriptstyle ?}}

\definecolor{darkblue}{rgb}{0.2, 0.2, 0.6}
\definecolor{darkgreen}{rgb}{0.0, 0.5, 0.0}

\newcommand{\M}{\mathcal{M}}
\newcommand{\phasespace}{\mathbb{G}}

\newcommand{\configspace}{\mathbb{Q}}
\newcommand{\momspace}{\mathbb{P}}

\newcommand{\realone}{\mathbb{R}}
\newcommand{\realpos}{\mathbb{R}^+}

\newcommand{\integerpos}{\mathbb{Z}^+}
\newcommand{\integernonneg}{\mathbb{Z}^+_0}
\DeclareMathAlphabet\mathbfcal{OMS}{cmsy}{b}{n}

\newcommand{\bGamma}{\mathbf{\Gamma}}

\newcommand{\bQ}{\mathbf{Q}}

\newcommand{\N}{\mathcal{N}}

\newcommand{\bP}{\mathbf{P}}

\newcommand{\E}{\mathcal{H}}
\newcommand{\K}{\mathcal{K}}

\newcommand{\energy}{\mathscr{E}}

\newcommand{\G}{\mathcal{G}}

\newcommand{\II}{\mathfrak{I}}
\newcommand{\tfree}{\tilde{\mathcal{F}}}

\newcommand{\free}{\mathcal{F}}
\newcommand{\partition}{\mathcal{Z}}

\newcommand{\R}{\mathcal{R}}
\newcommand{\I}{\mathcal{I}}
\newcommand{\bII}{\mathbfcal{I}}
\newcommand{\bvartheta}{\boldsymbol{\vartheta}}
\newcommand{\templim}{${T\to 0\;}$}
\newcommand{\bX}{\mathbf{X}}

\newcommand{\bh}{\mathbf{h}}

\newcommand{\DP}{\Delta_{\scriptscriptstyle \mathrm{P}}}
\newcommand{\DQ}{\Delta_{\scriptscriptstyle \mathrm{Q}}}
\newcommand{\SP}{\sigma_{\scriptscriptstyle \mathrm{P}}}
\newcommand{\SQ}{\sigma_{\scriptscriptstyle \mathrm{Q}}}

\newcommand{\rect}{\mathfrak{R}}
\newcommand{\ratio}{\mathfrak{r}}

\newcommand{\quotes}[1]{\text{`}#1\text{'}}
\newcommand{\result}{\mathfrak{M}}

\newcommand{\cover}{\mathcal{C}}
\newcommand{\pcover}{p^{\scriptscriptstyle (\cover)}}
\newcommand{\pcovers}[1]{p^{\scriptscriptstyle (\cover_{#1})}}
\newcommand{\hmedium}{h_{m}}
\begin{document}


\title{Derivation of Bose-Einstein statistics from the uncertainty principle}

\author{Paul Tangney}%
\email{p.tangney@imperial.ac.uk}
\affiliation{%
Department of Materials and Department of Physics, 
Imperial College London
}%

\date{\today}

\begin{abstract}
The microstate of any degree of freedom of any classical dynamical system
can be represented by a point in its two dimensional phase space. 
Since infinitely precise measurements are impossible,
a measurement can, at best, constrain the location of this point to a region 
of phase space whose area is finite. 
This paper explores the implications of assuming that this finite area
is bounded from below. I prove that if the same lower bound applied
to every degree of freedom of a sufficiently-cold classical dynamical system,
the distribution of the system's energy among its degrees of freedom 
would be a Bose-Einstein distribution.
\end{abstract}

\maketitle

\section{Introduction}
\label{section:introduction}
The development of quantum theory began with the discovery
that energy radiating from a  body at thermal equilibrium 
is not distributed among frequencies ($f$) as expected from (classical) statistical
mechanics~\cite{planck_1901,mehra,Gorroochurn}. 
The only ways found to derive the experimentally-observed distribution involved assuming that
either radiation itself, or the energy of an emitter of thermal radiation, was quantized
into indivisible amounts ${h f}$, where 
${h\approx \SI{6.6e-34}{\metre\squared\kilo\gram\per\second}}$ became
known as {\em Planck's constant}~\cite{planck_1901}.
The distribution of energy among frequencies that this quantization
implies became known as the {\em Bose-Einstein distribution}, in recognition
of the refinement and extension of Planck's work by Bose and Einstein~\cite{bose1924,einstein1924,einstein1925}.

The discrepancy between the observed spectrum of a hot object and the expected one
implied that the expectation was wrong. 
Planck's recognition that it could be resolved by assuming that light {\em emitters}
have quantized energies~\cite{planck_1901} led Einstein to the conclusion
that the energy of light itself is quantized~\cite{Einstein_1905}. Light quanta 
later became known as {\em photons}~\cite{Lewis_1926}.
Here I show that the discrepancy can be resolved without concluding that either 
light itself, or emitters of light, have quantized energies. 
It can be resolved by assuming the existence of a universal lower bound
on the precision to which the instantaneous 
microstate of any classically-evolving degree of freedom can
be measured or known.

The Bose-Einstein distribution is generally regarded as 
among the most significant deviations of quantum physics from classical physics, and 
among the characteristics by which bosons differ from fermions. However
the derivation of it presented here implies that any sufficiently-cold continuously-evolving
classical dynamical system would be \emph{observed} to obey Bose-Einstein statistics if the information
provided by observations and measurements of it was limited
by an uncertainty principle of the form
${\DQ\DP>\hq>0}$, where ${\DQ}$ and ${\DP}$ are the uncertainties in the values of 
the canonically-conjugate variables that specify a microstate of a single degree of freedom.
\vspace{-0.5cm}

\subsection{Assumptions}
\label{section:assumptions}
Consider an arbitrary isolated continuously-evolving deterministic
dynamical system, and let the instantaneous microstate
of one of its degrees of freedom (DOF) be specified by
a coordinate ${Q_t\in\configspace}$ and the 
coordinate's conjugate momentum, ${P_t\in\momspace}$, 
where ${\configspace\cong\realone}$ and ${\momspace\cong\realone}$
are the spaces of all possible coordinates and momenta, respectively.

Since the precision of any measurement is finite, the location of the point ${\Gamma_t\equiv(Q_t,P_t)}$ 
in the DOF's phase space, ${\phasespace\equiv\configspace\times\momspace\cong\realone\times\realone}$, 
cannot be known to infinite precision.
There is always some degree of uncertainty in the values of $Q_t$ and $P_t$.
Therefore the only state of \emph{certain} knowledge, as distinct from probabilistic knowledge,
that an observer could possess about the point ${\Gamma_t}$ is that it is
\emph{somewhere} in a specified finite-area subset of ${\phasespace}$.

For simplicity, and because only the most
accurate and precise measurements of the microstate 
are relevant to this work, let us assume that 
all measurements of ${Q_t}$ and ${P_t}$ 
result in the identifications of \emph{interval}
subsets of ${\configspace}$ and ${\momspace}$, respectively, 
which are certain to contain them.
Then all states of sufficiently-high certain knowledge 
about the location of ${\Gamma_t}$ in ${\phasespace}$
can be communicated as four values, $Q$, $P$, $\DQ$, and $\DP$, 
which specify a rectangular subset of ${\phasespace}$ 
with vertices ${(Q\pm\DQ/2,P\pm\DP/2)}$ that is known to contain ${(Q_t,P_t)}$.

The precisions, ${\DQ}$ and ${\DP}$, to which ${Q_t}$ and ${P_t}$
can be determined depend in part
on the microstate of the dynamical system at the time of measurement, and 
in part on how the measurement is performed.

\subsubsection{Uncertainty principle}
\label{section:uncertainty}
The unavoidably perturbative nature of the act of observation, and the
fact that it is impossible for an observer to possess an infinite 
amount of information, imply that ${\DQ\DP >0}$, but does not 
necessarily imply that the finite value of ${\DQ\DP}$ cannot be arbitrarily small.

However a finite universe contains a finite amount of information.
Therefore a lower bound on the value of ${\DQ\DP}$ must exist if the universe is finite.
For example, it is safe to say that the values of ${\DQ}$ and ${\DP}$ cannot be smaller 
than ${10^{-N_p}}$ in SI units, where ${N_p}$ is the number of particles in the universe.
Therefore it is safe to say that there never has been, and never will be, a measurement of a DOF's microstate 
which determined the location of the microstate in its phase space to within an  
area of less than ${10^{-2N_p}}$ in SI units.

This extreme example
demonstrates that ${\DQ\DP}$ would be bounded from below in a finite universe, and
its extremeness illustrates that larger universal lower
bounds on microstate precision must also exist (e.g., ${\DQ\DP> 10^{-N_p}}$ in SI units).
Only the largest of all universal lower bounds would be relevant to this work.

In an infinite universe, it is not immediately obvious that ${\DQ\DP}$ cannot be arbitrarily
small. However, uncertainty principles of the form ${\DQ\DP>\hq}$ arise in many contexts, 
and these uncertainty principles do not need 
to be universal (applicable in every possible context)
for the derivation presented in this work to imply that, at low $T$, the distribution of
a classical system's energy would appear to have the Bose-Einstein form 
in the context in which a particular uncertainty principle applies and is inviolable.

For example, Ref.~\onlinecite{macromicro} examines the relationship between macrostructure and
microstructure, where \emph{macrostructure} is the homogenized form that 
a microscopically-fluctuating classical field (the \emph{microstructure}) is observed to have 
on a much larger time and/or length scale (the \emph{macroscale}).
It is shown that an uncertainty principle of the form ${\DQ\DP>\hq}$ applies
at the macroscale when the probe used for all measurements is 
a macroscopic field (i.e., a homogenized microscopic field).
Therefore if the direct or indirect source of all empirical knowledge was
measurements with a macroscopic field, the energies of all 
sufficiently-cold classical systems would appear to be
Bose-Einstein distributed.

Another example would be a dynamical system that
was immersed in a bounded elastic medium.
The wavelengths and frequencies ($f$) of classical waves in a bounded uniform
medium are quantized, and the energy of a wave of amplitude $A$
can be expressed as ${\frac{1}{2}\gamma A^2 f^2}$, for some medium-dependent
constant ${\gamma}$. If $\Delta f$ was the frequency quantum, the smallest
energy difference between two waves, one of whose frequencies was $f$, would be
\begin{align*}
\Delta\energy &= \frac{1}{2}\gamma A^2\left(f+\Delta f\right)^2-\frac{1}{2}\gamma A^2f^2 
= \left(\gamma A^2\Delta f\right) f + \order{\Delta f^2}.
\end{align*}
Therefore if all of an observer's knowledge about the immersed object
had been communicated to them via the medium's waves, the smallest change in 
the energy of the object
that could be communicated to them by observing the change in energy of a 
wave of frequency ${f}$ 
would be ${\hmedium f}$,
where ${\hmedium=\gamma A^2 \Delta f}$.

I base the otherwise-classical derivation presented in this work on the following nonstandard 
and strong assumption: 
There exists a finite lower bound $\hq$ on the value of ${\DQ\DP}$,
and the same lower bound on microstate measurement precision applies to every observer and
to every DOF of every classical dynamical system.
In other words, I assume the existence of an uncertainty principle, ${\DQ\DP>\hq>0}$, that is \emph{universal}, meaning
valid in every possible context. 
However, as discussed, if an uncertainty principle has a restricted validity, the derivation shares the uncertainty 
principle's domain of validity.

For the purposes of this work I will assume that all measurements 
of a DOF's microstate are performed at the lower bound on microstate
precision, ${\DQ\DP=\hq}$. Therefore each measurement
of ${\Gamma_t}$ reveals that it is in a rectangle of area ${\hq}$,
centered at a point
${\Gamma}$, whose sides are parallel to the $\configspace$ and $\momspace$ axes.
I will denote such a rectangle by ${\rect(\Gamma,\ratio)}$, where
${\ratio\equiv\DQ/\DP=\DQ^2/\hq=\hq/\DP^2}$.

I will use the assumption ${\DQ\DP>\hq>0}$ to prove that, at thermal equilibrium, the distribution
of any classical dynamical system's energy among its DOFs is a
Bose-Einstein distribution in the low temperature (${T}$) limit, 
albeit with $\hq$ in place of Planck's constant, $h$.

\subsubsection{Low temperature limit}
\label{section:lowT}
The low $T$ limit is the
limit in which Bose-Einstein statistics apply within quantum mechanics. 
Both classically and quantum mechanically, the low $T$ limit is the weakly-interacting limit, 
and the Bose-Einstein distribution cannot be derived without assuming 
that interactions are weak enough to be approximated as absent for some purposes.

However it is important to clarify that assuming that an isolated physical system is in 
the low $T$ limit means assuming that interactions are arbitrarily weak, but finite.
It does not mean assuming that interactions are absent. 
This distinction is important because there would
not be any energy exchange between DOFs if interactions were absent. Therefore a
state of thermal equilibrium could never be reached, and it would not be meaningful to speak
of the physical system having a temperature.

Let us assume that each DOF's pair of canonically-conjugate
phase space coordinates, ${(Q_\eta,P_\eta)}$,
has been chosen such that, at any temperature,
the Hamiltonian can be expressed exactly as
\begin{align*}
\E\left(\{(Q_\eta,P_\eta)\}\right) = U\left(\{Q_\eta\}\right) + \sum_\eta \K_\eta\left(P_\eta\right),
\end{align*}
where $U$ is the potential energy and $\K_\eta$ is the kinetic energy
of DOF $\eta$.

Cooling the system brings its set of coordinates, ${\{Q_\eta\}}$, closer
to a local minimum of ${U}$. Therefore, by reducing $T$, ${\{Q_\eta\}}$ can be brought
arbitrarily close to a set, ${\{Q^{\text{min}}_\eta\}}$, 
at which the partial derivative ${\pdv*{U}{Q_\eta}}$ vanishes for
every $\eta$, and the second partial derivatives ${\pdv*[2]{U}{Q_\eta}{Q_\mu}}$
are all either zero or positive. Furthermore, it is always possible
to choose the set ${\{Q_\eta\}}$ such that the mixed derivatives, ${\pdv*[2]{U}{Q_\eta}{Q_{\mu\neq \eta}}}$, 
vanish at ${\{Q_\eta\}=\{Q_\eta^{\text{min}}\}}$.
Therefore it is always possible to express the potential energy as
\begin{align*}
U= U^{\text{min}} + \frac{1}{2}\sum_{\eta}\pdv[2]{U}{Q_\eta}\eval_{\{Q_\eta^{\text{min}}\}}\Delta Q_\eta^2 + \order{\Delta Q^3},
\end{align*}
where ${U^{\text{min}}\equiv U(\{Q^{\text{min}}_\eta\})}$ and ${\Delta Q_\eta\equiv Q_\eta-Q_\eta^{\text{min}}}$.

Reducing $T$ reduces the thermal averages of the ${\Delta Q_\eta}$'s and the standard deviation
of their fluctuations. Therefore, by cooling to a sufficiently low $T$, 
the terms of orders ${\Delta Q^3}$ and higher can be made negligible. This means that, by reducing $T$, the potential energy
can be approximated arbitrarily closely as
${U\approx U^{\text{min}}+\sum_\eta U_\eta}$, where ${U_\eta=U_\eta(Q_\eta)\propto\Delta Q_\eta^2}$.
Therefore assuming that a physical system is in the low $T$ limit
allows the Hamiltonian to be approximated as 
\begin{align}
\E\approx U^{\text{min}}+\sum_\eta \E_\eta(Q_\eta,P_\eta),
\label{eqn:total}
\end{align}
where ${\E_\eta(Q_\eta,P_\eta)\equiv U_\eta(Q_\eta)+\K_\eta(P_\eta)}$.

Since none of the terms on the right hand side of Eq.~\ref{eqn:total} depend on the phase space coordinates
of more than one DOF, if the derivatives ${\pdv*[2]{U}{Q_\eta}}\eval_{\{Q_\eta^{\text{min}}\}}$ are positive
the DOFs only exchange energy through the neglected terms 
in the potential energy. These terms can be made arbitrarily small by reducing $T$, so 
interactions between DOFs can be made arbitrarily weak by reducing $T$.

If the derivatives ${\pdv*[2]{U}{Q_\eta}}$ are all zero, then each ${\E_\eta}$ is
independent of ${Q_\eta}$, meaning that the system is gaseous. Therefore
the DOFs only exchange energy during rare and brief ``collisions,'' 
i.e., when the constant rates of change of 
two or more coordinates bring the set ${\{Q_\eta\}}$ into a region of the
configuration space ${\configspace}$ where $U$ is not independent of the ${Q_\eta}$'s.
When that happens, the coordinates either condense into a set of weakly-interacting oscillators, 
or cease interacting again. If they cease interacting, their kinetic energies after the collision
differ, in general, from their kinetic energies before the collision. If the
duration of each collision is comparable to the time between collisions, $T$ can be reduced until 
the former is a negligible fraction of the latter.

Regardless of whether a DOF becomes part of a set of weakly-interacting oscillators in the \templim~limit, 
or becomes an independent entity with constant potential energy in that limit, 
the assumption of a state of thermal equilibrium implies that energy \emph{is} exchanged - either slowly or rarely.
Therefore the equipartition theorem applies, which means that the time average of each DOF's energy is ${\frac{1}{2}k_B T}$, 
where ${k_B}$ is Boltzmann's constant.

\subsection{Outline of the derivation}
The uncertainty principle is the only non-standard assumption that I make to show that 
the energy of every  classical dynamical system is 
Bose-Einstein distributed in the \templim~limit.

To derive this result, I take an information theoretical approach to 
statistical mechanics that is very similar to the one introduced, or championed, by Jaynes~\cite{jaynes1,jaynes2}.
Jaynes' approach leans heavily on the work of Shannon~\cite{shannon}. 

There are three important steps in the derivation. The first step, which I discuss in detail in Sec.~\ref{section:unfalsifiable}, is to recognize that,
in the presence of uncertainty, the only empirically-unfalsifiable theories are statistical
theories, and that the only empirically-unfalsifiable statistical theories are
those in which uncertainty is maximised subject to the constraint that
everything that is known about the system is true.
I refer to the set of all known information pertaining to a physical system as the system's {\em macrostate}.

The second step, which I discuss in detail in 
Sec.~\ref{section:domain_quantization}, is to
recognize that when an uncertainty principle applies, 
the domains of empirically testable probability distributions are quantized.

The third step is to transform the coordinates ${(Q_\eta,P_\eta)}$ canonically, such that ${\E_\eta}$ is
transformed to a Hamiltonian with a particular form.

I will now outline the third step and explain why the derivation applies to \emph{every} sufficiently cold
classical dynamical system.

\subsubsection{Transforming the Hamiltonian of each degree of freedom to an affine form}
In Sec.~\ref{section:derivation} I will show that
when the uncertainty principle applies
there is no inconsistency between the Bose-Einstein distribution and the Maxwell-Boltzmann
distribution for a classical dynamical system: If the system is cold enough, 
the latter becomes the former under a canonical transformation of the set ${\{(Q_\eta,P_\eta)\}}$ 
of all phase space coordinates to a new set ${\{(X_\eta,Y_\eta)\}}$, which
transforms the Hamiltonian of each DOF from
${\E_\eta(Q_\eta,P_\eta) = U_\eta(Q_\eta)+\K_\eta(P_\eta)}$ to one of the form
\begin{align*}
\tilde{\E}=\tilde{\E}_0+\sum_\eta\tilde{\E}_\eta= \tilde{\E}_0+\sum_\eta \left[B_\eta + C_\eta X_\eta\right], 
\end{align*}
where ${\tilde{\E}_0}$ is constant, and
${B_\eta}$, ${C_\eta}$, and $Y_\eta$ are (approximately)
constants of the motion of DOF $\eta$.
Such a transformation is possible for \emph{every} sufficiently-cold classical dynamical system 
at thermal equilibrium because, as discussed in Sec.~\ref{section:lowT}, each DOF is either a
harmonic oscillator in that limit, or has a constant potential energy almost all of the time in that limit.

If the potential energy is constant, the only
energy that the DOF can exchange with other DOFs is its kinetic energy, ${\E_\eta=\K_\eta\propto P_\eta^2}$.
If $Q_\eta$ oscillates harmonically, the energy of DOF $\eta$ is proportional
to the square of its oscillation amplitude, i.e.,  ${\E_\eta\propto A_\eta^2}$. 
Therefore ${\E_\eta}$ has the same mathematical form in each case, and this quadratic
function of a single variable can be transformed canonically into an affine function of a single variable $X_\eta$
whose form is ${\tilde{\E}_\eta=B_\eta+C_\eta X_\eta}$~\cite{free_particle}.

For example, by transforming to \emph{action-angle coordinates} 
${(Q_\eta,P_\eta)\mapsto(\I_\eta,\theta_\eta)}$,
the Hamiltonian of a set of harmonic oscillators
is transformed from ${\E=\frac{1}{2}\sum_\eta \left[P_\eta^2 + \omega_\eta^2 Q_\eta^2\right]}$
to ${\tilde{\E}=\sum_{\eta}\I_\eta\omega_\eta}$, where ${\omega_\eta=\dot{\theta}_\eta}$
and the action ${\I_\eta}$ is a constant in the \templim limit of arbitrarily weak
interactions~\cite{landau,arnold,lanczos}.

\section{Unfalsifiable statistical models of deterministic systems}
\label{section:unfalsifiable}
The purpose of this section is to 
explain the concept of an {\em unfalsifiable statistical model} of a
classical Hamiltonian system.
An example of such a model is
the 19th century classical theory of thermodynamics.
Some readers may wish to skip to Sec.~\ref{section:derivation}, and to return
if or when they wish to scrutinise the logical foundations of the
derivation more carefully.

I begin by explaining what I mean by an unfalsifiable statistical model.
Then I explain my theoretical setup, before using this setup 
to derive the Maxwell-Boltzmann distribution. In Sec.~\ref{section:bose_einstein}
I show that, simply by changing the set of coordinates with which the microstate of
a set of oscillators or waves is specified, the Maxwell-Boltzmann distribution
becomes the Bose-Einstein distribution, albeit with an unknown constant
in place of Planck's constant.

To understand what I mean by an unfalsifiable statistical theory or model,
it is crucial to understand the difference between a {\em macrostate} and 
a {\em microstate}. 

\subsection{Macrostates and microstates}
A classical {\em microstate} is complete information about the state
of a deterministic system. It is a precise specification of 
the positions and momenta of all degrees of freedom of the system,
or the values of any variables from which these positions
and momenta could, in principle, be calculated.

A classical {\em microstructure} is complete information
about the structure of a deterministic system,
without any information about its rate of change with respect to time.

A {\em macrostate} $\M$ is simply a specification of
the domain of applicability of a particular unfalsifiable statistical
model.  A macrostate is a set of information specifying everything that
is known about the system to which the model applies. 
Because the model is statistical, it could only be falsified by a very 
large number of independent measurements. The macrostate is the complete list of everything that
the samples on which these measurements are performed are known to 
have in common. 
It is also the complete list of everything that is known about each individual sample, 
and which may significantly influence the final reported result of the measurement, assuming that
the uncertainty in this result is quantified correctly and reported with it.

\subsection{Examples} 
\subsubsection{Toy example}
As a very simple example, let us suppose that $\M$ contains the following information only: 

{\em There are three lockable boxes, coloured red, green, and blue, at least one of which is unlocked. 
A ball has been placed inside one of the unlocked boxes. If more than one box is unlocked, the box
into which the ball has been placed was chosen at random.}

Let us suppose that an experiment on a system meeting specification $\M$ 
consists of an experimentalist checking which box the ball is in.
Then, the only empirically-unfalsifiable statistical model of the experiment's results
would be a probability distribution that assigns a probability
of ${\frac{1}{3}}$ to the ball being in each box. Any other model could be falsified by statistics
from an arbitrarily large number of repetitions of the experiment performed
on independent realisations of system $\M$.

The fraction of times the ball would be found in each box would be ${\frac{1}{3}}$
even if different experiments
were performed with different boxes locked, as long as the choice of which boxes
were locked was made without bias, on average.

The model would be falsified by the empirical data if, say, the red box was chosen to be locked more 
frequently than the blue or green boxes.
However, if that occurred, it would not mean that the 
unfalsifiable model was defective, but that it was being applied to the wrong macrostate. 
After the bias was discovered and quantified it would form part of the specification of a new macrostate, ${\M'}$, 
and an unfalsifiable statistical theory of ${\M'}$ would be developed. Then, if no further
macrostate-modifying peculiarities were found, the set of all subsequent repetitions of the experiment
would produce data consistent with the unfalsifiable statistical theory of ${\M'}$.

\subsubsection{Realistic example} 
\label{section:realistic_example}
While considering a more complicated example, it may be useful to have an infinite set of independent laboratories
in mind. The equipment in each laboratory may be different, and different
methods of measurement may be used in each one, but all are capable of 
measuring whatever quantities the unfalsifiable statistical model
applies to. They are also capable of correcting their measurements for artefacts of the
particular sample-preparation and measurement techniques they are using, and of accurately quantifying 
uncertainties in the corrected values.

Then one can imagine asking each laboratory to measure, say, the bulk
modulus $B$ of diamond at a pressure of ${\SI{100}{\giga\pascal}}$
and a temperature of ${\SI{100}{\kelvin}}$. 
In this case, the statistical model would be a probability distribution, ${p(B)}$, for the
bulk modulus of an infinitely large crystal (to eliminate surface effects, which are sample-specific)
at precisely those values of pressure and temperature.

In general, each laboratory will prepare or acquire their sample of diamond in their own way, use a different
method of controlling and measuring temperature and pressure, and use
a different method of measuring $B$.
In addition to the quantified uncertainties in the measured  value of $B$, 
each independently-measured value will be influenced to some unquantified degree by {\em unknown unknowns}, i.e., 
unknown peculiarities of the sample, the apparatus, and the scientists performing the measurements and analysing
the data.
However, we will assume that this `data jitter' either averages out, when the data from all laboratories
is compiled, or is accounted for when comparing the compiled data to the statistical model.

If ${p(B)}$ was an unfalsifiable statistical model of $B$, it would be identical
to the distribution of measured values.
To derive or deduce an unfalsifiable distribution, one must carefully avoid making
any assumptions, either explicitly or implicitly, about the sample
or the measurement, apart from the information specified by the macrostate.
This means maximising one's ignorance of every other property of
a sample of diamond at ${(P,T)=(\SI{100}{\giga\pascal},\SI{100}{\kelvin})}$.
This is achieved by maximising the uncertainty in the value of $B$
that remains when its probability distribution, $p$, is known. 

To derive an unfalsifiable distribution for a given macrostate, one must
express the information specified by the macrostate
as mathematical constraints on $p$.
Then, under these information constraints, one must find the distribution $p$ for which the uncertainty in  
the value of $B$ is maximised. 
Maximising uncertainty eliminates bias and means that the information content of $p$
is the same as the information content of the distribution of measured values of $B$.
The differences between each distribution and a state of total ignorance is the same: it
is the information about the value of $B$ implied by the macrostate when no further information
is available.

In summary, elimination of bias, subject to the constraint that information $\M$ is true, guarantees that the resulting statistical model 
of the physical system defined by $\M$ is unfalsifiable: 
It guarantees that the model would agree with a statistical model calculated from a very large 
amount of experimental data pertaining to physical systems about which $\M$, and only $\M$, is known to be true.

\section{Probability domain quantization}
\label{section:domain_quantization}
The purpose of this section is to explain why one consequence of the uncertainty
principle is that the most informative statistically unfalsifiable probability distribution
for the location of a physical system's microstate in its
phase space $\phasespace$ is not a probability density function whose domain is $\phasespace$, 
but a probability mass function whose domain is a partition of ${\phasespace}$.
In other words, the uncertainty principle quantizes the 
domain of any empirically testable probability distribution for the location of a
classical dynamical system's microstate.

The derivations of the Maxwell-Boltzmann distribution in 
Sec.~\ref{section:derivation} and the Bose-Einstein distribution in Sec.~\ref{section:bose_einstein}
are reasonably self-contained, and reading this section is unnecessary to understand
the gist of these derivations.
However, skipping this section makes the derivations' logical
foundations appear simpler than they are.

It also makes it \emph{appear} that the derivations are built on 
an unjustified assumption: Namely, that an observer is capable of
determining which point ${\bGamma}$ on a lattice
in phase space the microstate $\bGamma_t$ of a physical system is closest to.
This section will make clear that such an assumption is not made. 
It is important that it is not made because, as I now explain,
the uncertainty principle, ${\DQ\DP>\hq}$, implies that
an observer would not be capable of such a determination. 
Therefore that assumption would be a false premise.
 
If ${\Gamma}$ and ${\Gamma+\Delta\Gamma}$ are
adjacent points of the lattice in the phase space $\phasespace$ of a single DOF, and if
${\N_{\Gamma}}$
and ${\N_{\Gamma+\Delta\Gamma}}$ are the
sets of points in $\phasespace$ that are closer to 
${\Gamma}$ and ${\Gamma+\Delta\Gamma}$, respectively, than
to any other points of the lattice, then ${\N_{\Gamma}}$ and 
${\N_{\Gamma+\Delta\Gamma}}$ share a border. 
The limit ${\hq}$ on
microstate measurement precision implies that it is impossible
for an observer to determine which side of their shared border the DOF's microstate ${\Gamma_t}$ is on.
Furthermore, as discussed in Sec.~\ref{section:assumptions}, the result of
a measurement of ${\Gamma_t}$ is the identification of an element $\Gamma$
of $\phasespace$ and a ratio ${\ratio=\DQ/\DP\in\realpos}$, such that
${\Gamma_t\in\rect(\Gamma,\ratio)}$. The value of $\Gamma$ is not restricted
to a point on a lattice, and the probability is zero that, by chance, it turns out to be one
of the points of a particular lattice, because the measure of a
lattice in ${\phasespace}$ is zero.
Therefore it is impossible for an observer to determine which 
point on a specific lattice ${\Gamma_t}$ is closest to.

The purpose of Sec.~\ref{section:probability_spaces} is to discuss
\emph{probability spaces} that are capable of satisfying
Kolmogorov's probability axioms~\cite{kolmogorov, borovkov,varadhan,jaynes03};
and, in particular, some
difficulties that arise when
deriving a probability distribution for the location 
of ${\Gamma_t}$ in $\phasespace$. It is the uncertainty 
principle that causes the difficulties, and which
forces us to confront certain subtleties in the definitions
of probability spaces.

In order to resolve the difficulties,
while ensuring empirical testability
of the probability distributions
that will be derived in Secs.~\ref{section:derivation}
and~\ref{section:bose_einstein}, 
a detail will be added in 
Sec.~\ref{section:statisticians} 
to the infinite set of 
measurements ($M$ measurements in the limit ${M\to\infty}$) performed
on independently prepared physical systems
that we imagined in Sec.~\ref{section:unfalsifiable}.

This detail is a filtration of the $M$ results of those
measurements: We will imagine defining an
infinite set ${\{\pcover\}}$ of different probability mass functions, $\pcover$, each of 
which is consistent with a different subset of the ${M\to\infty}$ 
measurements, and each of whose domains is a different
partition of $\phasespace$.
The introduction of this detail will clarify the true meaning of the
apparently-false premise on which 
the derivations in Secs.~\ref{section:derivation}
and~\ref{section:bose_einstein} are built.

A brief clarification is that we can imagine that
each function ${\pcover}$ is assigned to a different agent, or `statistician', 
and each probability mass function $p$ that is derived in later sections 
is the function ${\pcover}$ that has been assigned to one of those statisticians.
This construction allows us to imagine calculating
each distribution ${\pcover}$ in two ways: The first is from the
statistics gathered by the statistician to whom $\pcover$ has been assigned (`Statistician $\cover$').
The second is by using Jaynes' approach, as discussed in Sec.~\ref{section:unfalsifiable}, 
and as will be used in Sec.~\ref{section:derivation},
to theoretically derive the distribution that Statistician $\cover$ would
be unable to falsify.

\subsection{Probability spaces} 
\label{section:probability_spaces}
To develop a probabilistic description of the location of a microstate in its phase
space, we must construct one or more probability spaces, each of which satisfy
Kolmogorov's axioms of probability~\cite{jaynes03,kolmogorov,borovkov,varadhan}.
For simplicity, let us consider a physical system with a \emph{single} DOF, whose
microstate is ${\Gamma_t\in\phasespace}$.

A probability space ${(S,\Sigma,P)}$ consists of
a \emph{sample space} $S$, a \emph{$\sigma$-algebra} $\Sigma$, and
a \emph{probability measure}, ${P:\Sigma\to[0,1]}$.
The sample space $S$ is the set of all mutually-exclusive \emph{outcomes} or results of 
a trial or measurement, and the ${\sigma}$-algebra $\Sigma$ is the set of 
all events to which $P$ assigns probabilities. 

${\Sigma}$ is a \emph{cover} of $S$, meaning that it is a collection
of subsets whose union is the whole set. However, 
it is not necessary for the elements of $S$, which are the mutually
exclusive outcomes, to be elements of ${\Sigma}$. 
The only properties of $\Sigma$ that are required for
the probability space to satisfy the axioms of probability are that
it is
a set of subsets of $S$, which includes $S$ itself, and which is closed under countable
unions (${A_1,A_2,\cdots\in\Sigma\implies \bigcup_{i=1}^\infty A_i\in\Sigma}$),
closed under countable intersections (${\bigcap_{i=1}^\infty A_i\in\Sigma}$),
and closed under complements (${A\in\Sigma\implies S\setminus A\in\Sigma}$).

In Sec.~\ref{section:result_space}, we will consider the construction of a probability space for the outcome of
a measurement of a microstate, in order to show that constructing it is straightforward.

In Sec.~\ref{section:location_space} we will consider the construction of
a probability space for the location of ${\Gamma_t}$ in $\phasespace$
in order to show that the uncertainty principle makes an unbiased construction of a single probability
space impossible.
To avoid introducing bias, it is necessary to introduce an infinite number of probability spaces.

In Sec.~\ref{section:statisticians}, a logical 
construction will be outlined, which resolves some conceptual difficulties that
arise when describing the location of ${\Gamma_t}$ in $\phasespace$ with 
an infinite number of probability distributions.
This lays the logical foundations for
the derivations of the Maxwell-Boltzmann and Bose-Einstein distributions
presented 
in Sec.~\ref{section:boltzmann} and~\ref{section:bose_einstein}, respectively.

\subsubsection{Probability space for the outcome of a measurement of $\Gamma_t$}
\label{section:result_space}
The assumption made in
Sec.~\ref{section:assumptions} was 
that the outcome of an accurate and maximally precise
measurement of the location of ${\Gamma_t}$ in $\phasespace$ would be
the identification of an element ${(\Gamma,\ratio)}$ of 
\begin{align*}
\Omega\equiv\phasespace\times\realpos =\left\{(\Gamma,\ratio): \Gamma\in\phasespace,\;\ratio\in\realpos\right\},
\end{align*}
such that
${\Gamma_t\in\rect(\Gamma,\ratio)}$,  and with no constraints placed on the 
values of ${\Gamma\in\phasespace}$ and ${\ratio\in\realpos}$ that might be discovered.

Let ${\result:S\to\Omega}$ denote a random variable that maps the outcome $o$
of a measurement of ${\Gamma_t}$ to an element ${\result(o)}$
of $\Omega$; 
and let ${\quotes{\Gamma_t\in\rect(\Gamma,\ratio)}}$
represent the outcome of the measurement that $\result$ would
map to the point ${(\Gamma,\ratio)\in\Omega}$. 
Quotes are placed around ${\Gamma_t\in\rect(\Gamma,\ratio)}$ to
indicate that ${\quotes{\Gamma_t\in\rect(\Gamma,\ratio)}}$ 
represents the measurement outcome, which is a \emph{revelation} and a \emph{piece of information},
rather than the location of $\Gamma_t$ that the information revealed implies.

The sample space for the measurement outcome is
\begin{align*}
\quotes{\Omega}\equiv \left\{\quotes{\Gamma_t\in\rect(\Gamma,\ratio)}: (\Gamma,\ratio)\in\Omega\right\}.
\end{align*}
It is denoted by ${\quotes{\Omega}}$ because  its elements
are in one-to-one correspondence with elements of ${\Omega}$, and the
quotes indicate that its elements are revelations, rather than locations.

The elements of $\quotes{\Omega}$ are mutually exclusive 
because, assuming that
${\Gamma_1\neq\Gamma_2}$ and/or ${\ratio_1\neq\ratio_2}$,
the outcome of the measurement can be 
${\quotes{\Gamma_t\in\rect(\Gamma_1,\ratio_1)}}$
or
${\quotes{\Gamma_t\in\rect(\Gamma_2,\ratio_2)}}$, but not both.
It cannot be both because the result of each measurement is the
revelation of a \emph{single} imprecisely specified location. Since
${\rect(\Gamma_1,\ratio_1)\neq\rect(\Gamma_2,\ratio_2)}$,
${\quotes{\Gamma_t\in\rect(\Gamma_1,\ratio_1)}}$
and
${\quotes{\Gamma_t\in\rect(\Gamma_2,\ratio_2)}}$ are two different revelations, 
and both cannot occur.

The mutual exclusivity of elements of $\quotes{\Omega}$ makes it
straightforward to define a probability space ${(\quotes{\Omega},\wp(\quotes{\Omega}),P_o)}$
for the measurement outcome, whose ${\sigma}$-algebra
is the \emph{power set} ${\wp(\quotes{\Omega})}$ of ${\quotes{\Omega}}$. 
A probability density function ${\rho_o:\Omega\to\realpos}$ can also be defined
such that, for any ${A\subseteq\Omega}$, 
\begin{align*}
P_o(\{o\in \quotes{\Omega}:\result(o)\in A\}) = \int_A \dd{w} \rho_o(w).
\end{align*}
This is the
probability that the measurement discovers
that ${\Gamma_t\in\rect(\Gamma,\ratio)}$ for
some ${(\Gamma,\ratio)}$ in ${A \subset \Omega=\phasespace\times\realpos}$.
It is \emph{not} the probability that
${\Gamma_t}$ is in a particular subset of ${\phasespace}$.
However, 
in the limit ${\hq\to 0^+}$, we could define ${\Omega}$ as
${\phasespace}$ instead of ${\phasespace\times\realpos}$, in which case
it would become such a probability.

Therefore, were it not for the uncertainty principle, it would not
be necessary to draw attention to 
the distinction between the outcome ${o\in \quotes{\Omega}}$ of a measurement,
and the element ${\result(o)}$ of the measureable space $\Omega$ to which it is mapped by $\result$.
The uncertainty principle
makes discussing this distinction important, because it
is the location of $\Gamma_t$ that
we wish to model statistically, and we cannot directly use the range ${\Omega}$
of ${\result}$ as the domain of a probability distribution
for its location.
We will explore the reasons for this next.

\subsubsection{Probability space for the location of $\Gamma_t$ in $\phasespace$}
\label{section:location_space}
To understand why it is not straightforward to define a probability space
for the location of ${\Gamma_t}$, consider that, although the elements of $\quotes{\Omega}$
are mutually exclusive, the elements of the set
\begin{align*}
\rect(\Omega)\equiv\left\{\rect(w):w\in\Omega\right\},
\end{align*}
which is the set
of all imprecisely-specified locations of ${\Gamma_t}$
that the measurement might discover,
are not mutually exclusive locations.
They are not mutually exclusive
because 
${\rect(\Gamma_1,\ratio_1)}$ and ${\rect(\Gamma_2,\ratio_2)}$ might intersect.
If they intersected, ${\rect(\Gamma_1,\ratio_1)}$ and ${\rect(\Gamma_2,\ratio_2)}$
would not be mutually exclusive locations, but
${\quotes{\Gamma_t\in\rect(\Gamma_1,\ratio_1)}}$
and
${\quotes{\Gamma_t\in\rect(\Gamma_2,\ratio_2)}}$ would still
be mutually exclusive revelations because only one of them, at most, would be revealed.

The fact that the elements of ${\rect(\Omega)}$ 
are not mutually exclusive means that it cannot
be treated as a sample space for the purpose of building
a probability space. However, the problem is more serious than this:
${\rect(\Omega)}$ cannot even be a subset of a probability space's
$\sigma$-algebra, because
probabilities cannot be assigned to 
intersections of elements of ${\rect(\Omega)}$, and because
a $\sigma$-algebra must be closed under intersections of 
countable numbers of its elements.

For example, a probability cannot be assigned to the
event ${\Gamma_t\in\rect(\Gamma_1,\ratio_1)\cap\rect(\Gamma_2,\ratio_2)}$, 
despite the fact that there is an intuitively clear sense in which ${\Gamma_t\in\rect(\Gamma_1,\ratio_1)\cap\rect(\Gamma_2,\ratio_2)}$
is a possibility. It cannot be assigned a probability because whether or not 
this possibility has been realised is unknowable.
It is unknowable because the area of 
${\rect(\Gamma_1,\ratio_1)\cap\rect(\Gamma_2,\ratio_2)}$ is less than ${\hq}$, which means
that to know that ${\Gamma_t\in\rect(\Gamma_1,\ratio_1)\cap\rect(\Gamma_2,\ratio_2)}$
would imply a violation of the uncertainty principle.

Therefore, \emph{in the context of defining the ${\sigma}$-algebra of a probability space}, 
the probability
\begin{align*}
\Pr\left( \Gamma_t\in\rect(\Gamma_1,\ratio_1)\cap\rect(\Gamma_2,\ratio_2)\right) 
\end{align*}
is not a meaningful quantity.
The only probabilities that are meaningful in that context
are the probabilities of events that can be known to have occurred or to have not occurred.

One illustration of the problems that would arise if ${\Gamma_t\in\rect(\Gamma_1,\ratio_1)\cap\rect(\Gamma_2,\ratio_2)}$ was regarded
as an event that could be assigned a finite probability is the fact that
the probability measure that assigned the probability would be inconsistent with statistics
gathered from an infinite number of measurements: The fraction of the measurements
that would discover that event ${\Gamma_t\in\rect(\Gamma_1,\ratio_1)\cap\rect(\Gamma_2,\ratio_2)}$ 
had occurred would be zero.

Therefore a probability space whose probability measure
would be consistent with an infinite number of measurements must be built
from a sample space $\cover$ that is a 
cover of $\phasespace$ whose elements are mutually disjoint
subsets of $\phasespace$ of area no less than $\hq$.
I use the term \emph{disjoint} in the unconventional weaker sense
that sets $A$ and $B$ are disjoint if the measure ${\abs{A\cap B}}$ of their intersection is zero.
Therefore elements of $\cover$ may share boundaries.

Unfortunately, there are an infinite number of covers of $\phasespace$ that
meet these specifications. Therefore there are an infinite number of
probability spaces that could be built for the location of $\Gamma_t$ in $\phasespace$, 
and choosing any one of them as the statistical model that describes $\Gamma_t$ 
would be to introduce bias.
For example, in general, the expectation value,
\begin{align*}
\expval{O}_{\cover}\equiv \sum_{c\in \cover} \Pr(\Gamma_t\in c) \left(\frac{1}{\abs{c}}\int_c \dd{\Gamma} O(\Gamma)\right),
\end{align*}
depends on which cover $\cover$ is chosen,
where ${\abs{c}^{-1}\int_c \dd{\Gamma} O(\Gamma)}$ is the average on ${c\subset\phasespace}$ of some function
${O:\phasespace\to\realone}$, and ${\abs{c}}$ is the area of $c$. 

To avoid bias, we must define, or be aware of the existence of, an infinite number of probability spaces: 
There is one probability space, ${(\cover,\wp(\cover),P_\cover)}$, and one probability distribution, 
${\pcover:\cover\to[0,1]; c\mapsto \pcover(c)\equiv P_\cover(c)}$, 
for each cover $\cover$.

The next step is to understand how each element of the infinite set ${\{\pcover\}}$ of probability distributions
could, in principle, be validated or invalidated
by statistics from an infinite set of measurements of ${\Gamma_t}$, each of whose
outcomes is an element of ${\quotes{\Omega}}$. If it is not possible to imagine
calculating a distribution from statistics, rather than deriving it theoretically, 
it cannot be claimed that the theoretically derived distribution is empirically unfalsifiable.

\subsection{An infinitude of statisticians}
\label{section:statisticians}
Let us restrict attention to the most informative 
probability distributions possible. Therefore, let 
us disregard covers whose  elements are larger than necessary, and only consider 
probability distributions $\pcover$ whose domains are 
covers $\cover$ whose
elements all have areas of exactly ${\hq+\delta\hq}$. Let us also
take the  limit ${\delta\hq\to 0^+}$, so that ${\delta\hq}$ can be regarded
as both finite and arbitrarily small. Let ${\Lambda}$ denote the set
of all covers that meet these specifications.

Now let us assume that the results of the ${M\to \infty}$
measurements are distributed among
an infinite number of statisticians, such that there is exactly one statistician (`Statistician $\cover$') for each
${\cover\in \Lambda}$.
Then let us imagine that each measurement whose
outcome is ${\quotes{\Gamma_t\in\rect(\Gamma,\ratio)}}$
is communicated to all of the 
statisticians whose covers 
contain an element of which ${\rect(\Gamma,\ratio)}$ is a subset, 
and is not communicated to the rest of the statisticians.

Clearly, every measurement of ${\Gamma_t}$ determines that  ${\Gamma_t\in\phasespace}$. Therefore, 
by definition of a \emph{cover}, every measurement determines that $\Gamma_t$ is in
\emph{some} element of \emph{every} statistician's cover. 
However, we are supposing that each statistician learns the result of each
measurement of $\Gamma_t$ if and only if the measurement has
determined \emph{which} element of their cover contains ${\Gamma_t}$.
This can only be the case if the set ${\rect(\Gamma,\ratio)}$ that
the measurement discovers $\Gamma_t$ to be in is a subset
of an element of their cover.

For each element ${c}$ of $\cover$, Statistician
${\cover}$ calculates the fraction, ${\pcover(c)}$,
of the total number ${M_\cover<M}$ of measurements whose outcomes they are privy to,
for which ${\Gamma_t\in c}$.
Therefore, in the limit ${M\to\infty\implies M_\cover\to\infty}$, Statistician~$\cover$
calculates a probability distribution ${\pcover}$, 
whose domain is $\cover$.

The next question to address is the following: If one of the $M$ measurements
was chosen at random, is ${\pcover(c)}$ 
the probability that $c$ contains
the microstate of the sample being measured in that measurement?
In other words, is ${\Pr(\Gamma_t\in c)=\pcover(c)}$?\\

The first thing to note is that ${\Pr(\Gamma_t\in c)}$ is an unknowable probability, 
for the same reason that, in general, ${\Gamma_t\in c}$ is an untestable proposition:
${\mathcal{P}_{\textrm{certain}}\equiv P_o(\{\quotes{\Gamma_t\in c}\})}$  is 
the fraction of the $M$ measurements
in which it is known that ${\Gamma_t\in c}$, and
\begin{align*}
\mathcal{P}_{\textrm{possible}}\equiv P_o\left(\left\{o\in\quotes{\Omega}: \rect(\result(o))\cap c\neq \emptyset\right\}\right)
\end{align*}
is the fraction of the $M$ measurements in which it is known that ${\Gamma_t\in c}$ is
possible. However it is impossible to know the fraction of
the $M$ measurements for which ${\Gamma_t\in c}$, for reasons discussed
in the introduction to Sec.~\ref{section:domain_quantization}: the uncertainty
principle implies that it can never be known which side of a border 
between elements of $\cover$
$\Gamma_t$ is on. Furthermore, because ${\quotes{\Omega}}$ has an infinite
number of elements, the ratio ${\mathcal{P}_{\textrm{certain}}/\mathcal{P}_{\textrm{possible}}}$
vanishes.

To shed more light on the empirically-unanswerable question of
whether ${\Pr(\Gamma_t\in c)=\pcover(c)}$,
let us consider the possibility that 
\begin{align*}
\pcovers{1}(c)\neq\pcovers{2}(c),
\end{align*}
for two covers
${\cover_1,\cover_2\in\Lambda}$ that both contain $c$.
By construction, every time the measurement outcome is ${\quotes{\Gamma_t\in c}}$, 
this outcome is revealed to both Statistician~${\cover_1}$ and Statistician~${\cover_2}$.
Therefore the numbers of times that these statisticians
learn that ${\Gamma_t\in c}$ are the same. Let us denote that number by $m_c$.
Therefore 
\begin{align*}
\pcovers{1}(c)\equiv \frac{m_c}{M_{\cover_1}}\neq\frac{m_c}{M_{\cover_2}}\equiv \pcovers{2}(c)
\end{align*}
would imply that, 
even in the ${M\to\infty}$ limit,
${M_{\cover_1}\neq M_{\cover_2}}$.
Therefore 
\begin{align*}
M_{\cover_1\setminus \{c\}} = M_{\cover_1}-m_c \neq M_{\cover_2}-m_c= M_{\cover_2\setminus \{c\}},
\end{align*}
where, for example, ${M_{\cover_1\setminus \{c\}}}$ is the number of times 
that it has been revealed to Statistician~${\cover_1}$ that ${\Gamma_t}$ is
in an element of ${\cover_1}$ that is not $c$.
This implies that
\begin{align*}
\sum_{c'\in\cover_1\setminus\{c\}}\pcovers{1}(c')
\neq
\sum_{c'\in\cover_2\setminus\{c\}}\pcovers{2}(c').
\end{align*}
Therefore, since ${\cover_1\setminus\{c\}}$ and ${\cover_2\setminus\{c\}}$
are both covers of ${\phasespace\setminus c}$, 
${\pcovers{1}(c)\neq\pcovers{2}(c)}$
would imply that the fraction of the $M$ measurements that
discover that ${\Gamma_t}$ is in a particular
subset of ${\phasespace}$ of dimensions ${\DQ\times\DP=\hq+\delta\hq}$, 
would not be determined solely by the fraction of the $M$ measured
samples for which ${\Gamma_t}$ is in that subset

In other words (and for clarity I will use the 
unjustifiable and unphysical assumption that
it is possible to know the
probabilities ${\{\Pr(\Gamma_t\in g): g\subset \phasespace\}}$),
${\pcovers{1}(c)\neq\pcovers{2}(c)}$ would imply
that there does not exist a constant $K$ such 
that ${P_o(\quotes{\Gamma_t\in\rect(\Gamma,\ratio)})=K\Pr(\Gamma_t\in\rect(\Gamma,\ratio))}$
for all ${(\Gamma,\ratio)\in\Omega}$.

Not only can we not rule out the possibility that $K$ is not constant,
it would be surprising if it were constant:
It was mentioned in Sec.~\ref{section:assumptions} that
the measurement precisions ${\DQ}$ and ${\DP}$ depend in part on
$\Gamma_t$ and in part on how the measurement of $\Gamma_t$ is
performed. Therefore, if the measurement 
outcome is ${\quotes{\Gamma_t\in\rect(\Gamma,\ratio)}}$, 
the location of $\Gamma_t$ in $\phasespace$ 
has played a part in determining $\ratio$, in general.
The fact that it has also played a part in determining $\Gamma$
is obvious.

However the dependence of $K$ on the microstate
of a physical system implies that $K$ depends on the system's Hamiltonian, 
which implies that it depends on what 
the physical system is. In other words, this dependence
cannot be a universal limitation on the act
of measuring a DOF's microstate. 

Therefore, instead of abandoning the prospect of
devising a universally-applicable statistical model, such
as Bose-Einstein statistics,
this dependence should be treated as one of the pecularities
of individual physical systems, or methods of measurement,
that were discussed in Sec.~\ref{section:realistic_example}, 
and whose effects on statistics must be accounted for before
those statistics can be compared with the predictions
of universally-applicable statistical models.
When deriving a statistical model that is universally applicable,
it is not only reasonable to assume that $K$
is the same for every ${(\Gamma,\ratio)\in\Omega}$,
making that assumption appears to be unavoidable.

In other words, while bearing in mind that
${\pcover(c)\propto P_o(\quotes{\Gamma_t\in c})\propto\Pr(\Gamma_t\in c)}$
is an empirically untestable proposition, let us use  it as 
a rough approximation to a more nuanced and precise
interpretation of ${\pcover(c)}$.
Then, so that we can derive a universally-applicable statistical model, 
we purposely neglect pecularities of individual physical systems, and samples of
those systems, because this is the only way to derive a model that is generally applicable. 
This entails assuming that the fraction of the ${M_\cover}$ measurements
revealed to Statistician~$\cover$ for which ${\Gamma_t\in c}$ 
equals the fraction of all $M$ measurements for which ${\Gamma_t\in c}$.

\subsubsection{Justification of a working assumption used in the derivations}
As discussed above, ${\pcover(c)=\Pr(\Gamma_t\in c)}$ is
an empirically untestable proposition, but is also the only reasonable 
assumption to make when deriving a generally-applicable unfalsifiable
probability distribution. It is equivalent to the assumption that
the number of measurements whose outcome is ${\quotes{\Gamma_t\in\rect(\Gamma,\ratio)}}$
is proportional to the number of measurements in which ${\Gamma_t\in\rect(\Gamma,\ratio)}$, 
with the same constant of proportionality for every $\Gamma$ and every $\ratio$.

Under the assumption that ${\pcover(c)=\Pr(\Gamma_t\in c)}$, 
we can justify the working assumption that it is possible 
to determine which element of
cover $\cover$ contains $\Gamma_t$ as follows:
From the perspective of Statistician~$\cover$, 
the revelation of a measurement outcome to them can be regarded
as their `measurement' of ${\Gamma_t}$. Therefore, from 
their perspective, each of their measurements determines which 
element of ${\cover}$ contains
${\Gamma_t}$.

Then we can imagine that
Statistician~$\cover$ calculates $\pcover(c)$
from the results of their `measurements', and that
if we are told the macrostate ${\M}$ that
defines the measurements, we can theoretically derive
a probability distribution whose domain is $\cover$, 
and which agrees perfectly with $\pcover(c)$, 
by eliminating all bias subject to the constraint
that information $\M$ is true.

Each of the distributions derived in Sec.~\ref{section:derivation}
and Sec.~\ref{section:bose_einstein} can be interpreted
as this theoretically-derived statistically-unfalsifiable probability
distribution, where the statistics that fail to falsify it
are those gathered by Statistician~$\cover$.

\section{Derivation of an unfalsifiable energy distribution}
\label{section:derivation}
Section~\ref{section:setup} presents a brief summary of the theoretical setup that is used 
in Sec.~\ref{section:boltzmann} and Sec.~\ref{section:bose_einstein} to
derive the Maxwell-Boltzmann distribution 
and the Bose-Einstein distribution, respectively.

It is assumed that it is possible for a measurement
to determine which element of a cover of a DOF's 
phase space, comprising disjoint subsets of area $\hq$, contains
the DOF's microstate. Although this assumption is not compatible
with the uncertainty principle discussed in Sec.~\ref{section:uncertainty}, 
its use in derivations as a \emph{working assumption} was 
justified in Sec.~\ref{section:statisticians}.

\subsection{Theoretical setup}
\label{section:setup}

Consider an arbitrary continuously-evolving deterministic system
whose microstate can be specified by 
${\bGamma\equiv(\bP,\bQ)}$,
where ${\bQ\equiv (Q_{1},Q_{2}\cdots)}$ 
is some set of generalized coordinates 
and
${\bP\equiv (P_{1},P_{2}\cdots)}$, where
${P_\eta}$ is
the momentum conjugate to ${Q_\eta}$.
In this coordinate system, 
let ${\E(\bGamma)}$ denote the system's Hamiltonian, 
and, as before, ${\phasespace\equiv\configspace\times\momspace\ni\bGamma}$, ${\configspace\ni\bQ}$, and ${\momspace\ni\bP}$
denote the system's phase space, configuration space, and momentum space, respectively.

Let us begin by partitioning $\phasespace$ 
into nonoverlapping subsets of equal measure (phase space `volume') as follows:
We choose a countable set $\G$ of evenly-spaced
points (microstates) in $\phasespace$ and define a neighbourhood ${\N_\bGamma\subset\phasespace}$
of each point ${\bGamma\in\G}$ such that ${\phasespace = \bigcup_{\bGamma\in\G}\N_\bGamma}$,
and such that, if ${\bGamma,\bGamma'\in\G}$ are any two different points (${\bGamma\neq\bGamma'}$), 
then
${\abs{\N_\bGamma\cap\N_{\bGamma'}}=0}$ and ${\abs{\N_\bGamma}=\abs{\N_{\bGamma'}}}$,
where
${\abs{\N_\bGamma}}$ denotes the measure of ${\N_\bGamma}$ in ${\phasespace}$.
For simplicity, let us assume that if ${\bGamma_t\in\N_\bGamma}$, then $\bGamma_t$
is closer to ${\bGamma}$ than to any other element of $\G$.
Therefore the interior of ${\N_\bGamma}$ is the set of all points in $\phasespace$ that
are closer to $\bGamma$ than to any other element of $\G$.

Now let  ${p_\bGamma}$, where ${\bGamma\in\G}$, 
denote the probability, ${\Pr(\bGamma_t\in\N_\bGamma)}$,
that $\bGamma_t$ is within ${\N_\bGamma}$.
The probability distribution for the point ${\bGamma}$ that 
identifies the region $\N_\bGamma$ containing ${\bGamma_t}$ is
${p:\G\to [0,1]; \bGamma\mapsto p_\bGamma}$.

Now let us suppose, momentarily, that $\bGamma_t$ is known to be in region $\N_\bGamma$, 
and that $\N_\bGamma$ is partitioned
into ${W_\bGamma}$ nonoverlapping subsets of equal measure ${v\equiv\abs{\N_\bGamma}/W_\bGamma}$.
Then, as Shannon demonstrated~\cite{shannon}, we can quantify the amount of information that 
must be revealed to determine which of these subsets ${\bGamma_t}$ is in by
${\log W_\bGamma = \log \abs{\N_\bGamma}-\log v}$. \\
In the limit ${W_\bGamma\to \infty,\; v\to 0}$, the quantity of information required
becomes infinite.  However, as discussed in Sec.~\ref{section:uncertainty},
we are assuming that $v$ has a lower bound, which means that
${W_\bGamma}$ has an upper bound. 

Without losing generality, let us assume that these bounds are $\abs{\N_\bGamma}$ and $1$, respectively. In other words, 
let us assume that when we originally partitioned $\phasespace$, we chose the 
set $\G$ such that the following is true:

{\em
Given any microstate ${\bGamma\in\G}$, and
any microstate ${\bGamma'\in\phasespace}$, which is closer to ${\bGamma}$ than to any other element of $\G$, it is theoretically possible
to distinguish between ${\bGamma'}$ and any element of ${\G\setminus\{\bGamma\}}$ by empirical means; and it is impossible
to distinguish between ${\bGamma'}$ and ${\bGamma}$ by empirical means}.

I will refer to $\G$ as a {\em maximal set of mutually-distinguishable microstates}; I will refer 
to a sampling of $\phasespace$ with such a set as a {\em maximal sampling}; and I will use ${\bh\equiv \abs{\N_\bGamma}}$ to denote
the measure of each neighbourhood ${\N_\bGamma}$ in a maximal sampling of phase space.

\subsection{Maxwell-Boltzmann statistics}
\label{section:boltzmann}
This section draws heavily from the works of Jaynes~\cite{jaynes1} and Shannon~\cite{shannon}.

Let us add the assumption that we know that the
expectation value of the system's energy is $\energy$.
For example, the system might be a classical crystal whose
average energy is determined by a heat bath to which it is coupled.

The system's state of thermal equilibrium can be defined as the probability distribution $p$ 
that maximises the Shannon entropy~\cite{shannon},
subject to the constraint that the Hamiltonian's expectation value,
\begin{align*}
\expval{\E}[p]\equiv \sum_{\bGamma\in\G} p_\bGamma\E(\bGamma),
\end{align*}
is equal to $\energy$, 
and subject to the normalization constraint ${\sum_{\bGamma\in\G} p_\bGamma = 1}$.
The Shannon entropy is
\begin{align}
\expval{S}[p] &\equiv \sum_{\bGamma\in\G}  p_\bGamma \II(p_\bGamma),
\label{eqn:entropy}
\end{align}
where 
${
\II(p_\bGamma) 
\equiv-\log  p_\bGamma}$
is the Shannon information~\cite{shannon} of
$p$ at $\bGamma$. From now on it will be implicit that ${\sum_\bGamma}$ means ${\sum_{\bGamma\in\G}}$.

The Shannon information, ${\II(p_\bGamma)}$,
quantifies how much would be learned, meaning by how much would the uncertainty in the location of $\bGamma_t$ reduce, 
if it was discovered that ${\bGamma_t\in\N_\bGamma}$.
The functions ${k\II(p_\bGamma)}$, for any ${k\in\realpos}$, are
the only functions that satisfy the following three conditions: (i) they would
vanish if it was known that ${\bGamma_t}$ was in ${\N_\bGamma}$
prior to `discovering' it there, i.e., 
if ${p_\bGamma=1}$; (ii) they increase as the discovery that ${\bGamma_t\in\N_\bGamma}$ becomes more surprising, i.e.,
as ${p_\bGamma}$ decreases; and (iii) they are additive. 
Additivity means that if, for example, it was discovered
that ${\bGamma_t}$ was in ${\N_\bGamma}$ 
and that the microstate ${\bGamma'_t}$ of another independently-prepared
and statistically independent system was in
${\N_{\bGamma'}}$, the 
quantity of information about the locations of 
${\bGamma_t}$ and ${\bGamma'_t}$ that was unknown would
decrease by ${\II(p_\bGamma)+\II(p_{\bGamma'})}$.

Any probability distribution, $p$, is a state of knowledge that an observer
could be in. The Shannon information, ${\II(p_\bGamma)}$, of $p_\bGamma$, 
quantifies the information that would be revealed by the discovery
that ${\bGamma_t\in \N_\bGamma}$, and the Shannon entropy is
the expectation value of the quantity of information
that would be revealed by
discovering which point $\bGamma$ in the maximal set of mutually-distinguishable
microstates $\G$ the true microstate $\bGamma_t$ is closest to.
Therefore ${\expval{S}[p]}$ quantifies the incompleteness of distribution $p$, as a state
of knowledge, when the identity of the element of $\G$
that is closest to ${\bGamma_t}$ is regarded as
complete knowledge. 

Whether or not ${\expval{S}[p]}$ 
is satisfactory as a quantification of
uncertainty in all contexts is probably irrelevant in the present context, because
we will be maximising its value subject to the stated
contraints. Therefore what is relevant is that it increases
monotonically as the location of $\bGamma_t$ in $\phasespace$
becomes more uncertain.

We can express the stationarity of ${\expval{S}[p]}$ subject to constraints ${\expval{\E}[p]=\energy}$
and ${\sum_{\bGamma} p_\bGamma=1}$
as 
\begin{align*}
\delta\left\{ \expval{S}[p] - \beta\left(\expval{\E}[p]-\energy\right)
-\beta\lambda\left(\sum_{\bGamma} p_\bGamma -1\right)\right\} =  0,
\end{align*}
where $\beta$ and ${\beta\lambda}$ are Lagrange multipliers.
If we divide across by ${-\beta}$ and define the constant ${T\equiv (k_B\beta)^{-1}}$, where
${k_B}$ is the Boltzmann constant, 
this can be expressed as ${\delta \left(\tfree[p]+\lambda\sum_{\bGamma}p_\bGamma\right) = 0}$, 
where ${\tfree[p]\equiv\expval{\E}[p]-k_B T\expval{S}[p]}$.
By taking a partial derivative of ${\tfree[p]+\lambda\sum_{\bGamma}p_\bGamma}$
with respect to $p_\bGamma$ and setting
it equal to zero, we find that
\begin{align}
p_\bGamma & = 
e^{-(\E(\bGamma)-\free)/k_B T}
 = \partition^{-1} e^{-\E(\bGamma)/k_B T},
\label{eqn:boltzmann}
\end{align}
where ${\partition\equiv \exp\left(-\free/k_B T\right)}$ is known 
as the {\em partition function} and 
we refer to
the quantity ${\free = -k_B T\log \partition}$, which 
is the value taken by ${\tfree[p]}$ when it is stationary
with respect to normalization-preserving variations of $p$, as the {\em free energy}.

Equation~\ref{eqn:boltzmann} is the familiar Maxwell-Boltzmann distribution
and $T$ is the temperature.
The derivation of Eq.~\ref{eqn:boltzmann} is a derivation, based on the
premises that precede it and those stated within it, of the only
empirically-unfalsifiable probability distribution for the true microstate.
It is unfalsifiable because it explicitly rejects bias by maximising uncertainty subject 
to one physical constraint, which is 
the only thing that we know about the state of the system; namely, that a heat bath
ensures that its average energy is $\energy$.

As discussed in Sec.~\ref{section:unfalsifiable}, the absence of bias guarantees us 
that if we had enough independent replicas of the physical system, and if the only
thing we knew about each one was that its average energy was $\energy$,
and if we could determine by measurement which element ${\N_\bGamma}$ of the phase space partition the microstate of each one was in, 
the fraction of those whose microstate was in ${\N_\bGamma}$
would be ${p_\bGamma = e^{-\beta\E(\bGamma)}/\partition}$.

Now let us make the simplifying assumption under which
the Bose-Einstein distribution is
valid within quantum mechanics: The total energy
is a sum of the energies of \emph{independent}
DOFs.  Within quantum mechanics these DOFs are often interpreted as particles.

With the Hamiltonian of DOF $\eta$ denoted by  ${\E_\eta(\Gamma_\eta)}$,  where ${\Gamma_\eta\equiv(Q_\eta,P_\eta)}$,
we can express the Hamiltonian
of the set of all  DOFs as
\begin{align}
\E(\bGamma) = \sum_\eta \E_\eta(\Gamma_\eta),
\end{align}
and we can express the partition function as
\begin{align}
\partition&\equiv  \sum_{\bGamma} e^{-\beta\E(\bGamma)} 
= \sum_{\bGamma} \prod_{\eta} e^{-\beta\E_{\eta}(\Gamma_\eta)} 
\label{eqn:partition_sum}
\end{align}
where the product ${\prod_\eta}$ is over all DOFs.

Now let us choose the maximal set of mutually-distinguishable microstates, $\G$, 
to be a lattice, which is the direct product ${\prod^\times_\eta \G_\eta}$, 
where ${\G_{\eta}}$ is both a two dimensional lattice and a maximal
set of mutually-distinguishable points in the phase
space ${\phasespace_{\eta}}$ of DOF ${\eta}$. 
The area of the non-overlapping neighbourhoods ${\N_{\Gamma_\eta}}$
of ${\Gamma_\eta}$ whose union is ${\phasespace_\eta}$ is 
${\hq\equiv \abs{\N_{\Gamma_\eta}}=\Delta Q_{\eta}\Delta P_{\eta}=\DQ\DP}$, 
where ${\frac{1}{2}\Delta P_\eta}$ is the smallest difference
in momentum ${P_\eta}$ between mutually-distinguishable microstates
of $\eta$ with the same coordinate; and ${\frac{1}{2}\Delta Q_\eta}$ is the smallest
difference in coordinate ${Q_\eta}$ between mutually-distinguishable microstates
with the same momentum.

These choices and definitions allow us to swap the order of the sum and the product in Eq.~\ref{eqn:partition_sum}, 
thereby expressing it as
${\partition=\prod_{\eta} \partition_{\eta}}$, where
\begin{align}
\partition_{\eta}\equiv 
\sum_{\Gamma_{\eta}} e^{-\beta\E_{\eta}(\Gamma_{\eta})},
\label{eqn:mode_partition}
\end{align}
and where ${\sum_{\Gamma_\eta}}$ denotes ${\sum_{\Gamma_\eta\in\G_\eta}}$.
If we know the partition function $\partition_{\eta}$ of each DOF ${\eta}$, we can calculate the partition
function $\partition$ of the system as a whole. 

In Sec.~\ref{section:bose_einstein} we will explore other ways to calculate $\partition$
by transforming away from ${(\bP,\bQ)}$ and ${(P_\eta,Q_\eta)}$ to different
sets of variables. To avoid a proliferation of new symbols, I will recycle the
symbols ${\phasespace}$, ${\phasespace_{\eta}}$,
${\E}$, ${\E_{\eta}}$, ${\bGamma}$, 
${\Gamma_{\eta}}$, $\N_\bGamma$, ${\G}$, ${\G_{\eta}}$, $\bh$, $p_\bGamma$, and $p$.
They will have the same meanings in the new coordinates as they do for coordinates ${(\bP,\bQ)}$.

\section{Bose-Einstein statistics} 
\label{section:bose_einstein}
I will now derive the Bose-Einstein distribution for a classical
system of non-interacting oscillators or standing waves.
Then I will briefly discuss how the derivation can be
generalized to other kinds of physical systems.

\begin{center}
\begin{figure*}[!]
\includegraphics[width=.9\textwidth]{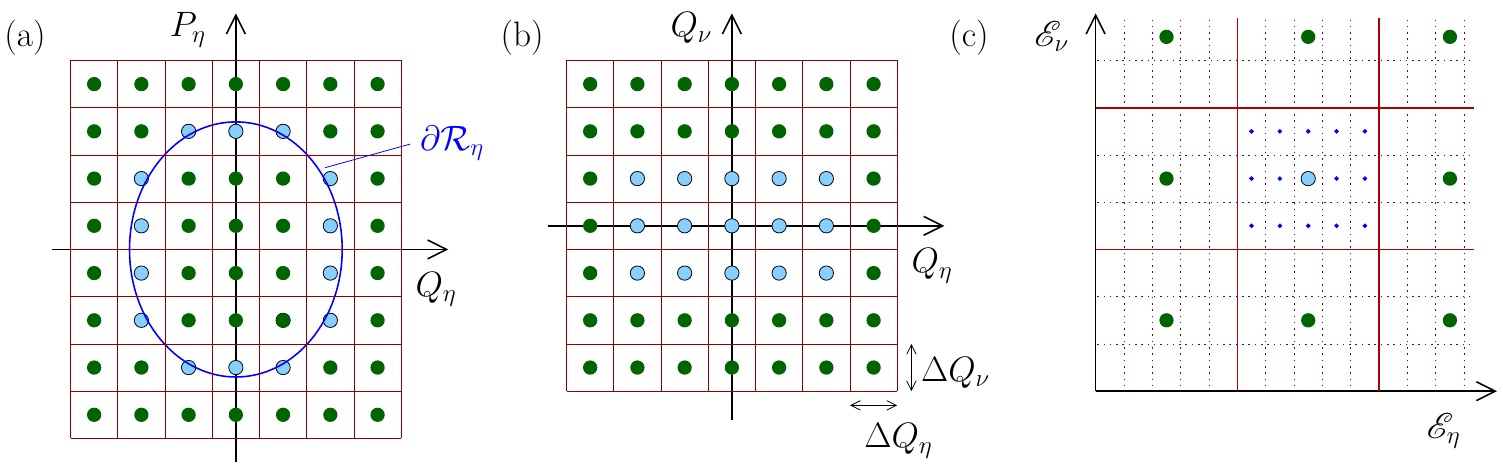}
\caption{(a) A portion of the phase space $\phasespace_{\eta}$ 
of mode ${\eta}$. The continuous blue ellipse, ${\partial\R_{\eta}}$, is a particular constant-energy path 
that the oscillation follows when it is decoupled from other modes.
The set of pale blue and green spots is a maximal set of mutually-distinguishable
microstates, $\Gamma_{\eta}$, of mode ${\eta}$. In 
statistical models of the mode's microstates, each spot represents
all points within its rectangular neighbourhood.
(b) A portion of the {\em microstructure} space of modes ${\eta}$ and ${\nu}$. The spots
belong to a maximal set of mutually distinguishable points and represent the rectangular regions
they inhabit. The pale blue spots mark regions visited during the motion of the modes, assuming that
their energies, ${\energy_{\eta}}$ and ${\energy_{\nu}}$, 
are constant and that neither of their frequencies, $\omega_{\eta}$ and $\omega_{\nu}$, is
an integer multiple of the other.
Each of the $15$ pale blue spots represents the four points
${(Q_{\eta},Q_{\nu},P_\eta,P_\nu)=
\left(Q_\eta,Q_\nu,
\pm\sqrt{2\energy_{\eta}-\omega_{\eta}^2Q_{\eta}^2},
\pm\sqrt{2\energy_{\nu}-\omega_{\nu}^2Q_{\nu}^2}\right)}$ in their joint phase
space ${\phasespace_{\eta}\times\phasespace_{\nu}}$.
(c) The pale blue spot is the energy of the trajectory represented by pale blue spots
in panels (a) and (b). We cannot calculate the partition 
function of modes ${\eta}$ and ${\nu}$ as
${\partition_{\eta}\partition_{\nu}
=\sum_{\energy_{\eta}}\sum_{\energy_{\nu}} e^{-\beta\left(\energy_{\eta}+\energy_{\nu}\right)}}$
if the double summation is over a square grid in ${(\energy_{\eta},\energy_{\nu})}$-space.
The numbers of energies sampled along each axis are only in the same ratio as the numbers of mutually-distinguishable mode coordinates
along each axis in ${(Q_{\eta},Q_{\nu})}$-space, 
and the numbers of mutually-distinguishable points in ${\phasespace_{\eta}}$ and ${\phasespace_{\nu}}$, if the spacings
of sampled values along the mode's energy axes are
their frequencies times the same constant. 
}
\label{fig:one}
\end{figure*}
\end{center}

\subsection{Oscillators and standing waves}
\label{section:oscillators}
As discussed in Sec.~\ref{section:lowT}, 
if the potential energy of a classical dynamical system
is a smooth function ${U(\bQ)}$ of its microstructure ${\bQ}$, the
system can be brought arbitrarily close to a minimum of its potential
energy, ${\bQ^{\text{min}}}$, by cooling it slowly.
Once ${\norm{\bQ-\bQ^{\text{min}}}}$ is small enough,
lowering its temperature further brings its dynamics 
closer to a superposition of harmonic oscillations
of the normal modes of its stable equilibrium structure, ${\bQ^{\text{min}}}$.
For example, a set of mutually-attractive particles
would condense into a stable vibrating cluster when cooled.
The normal modes of a finite crystal or a continuous bounded medium are standing waves, 
so their dynamics become superpositions of standing waves when they are cold enough.

If we specify the microstructure by the set of displacements from
mechanical equilibrium along the normal mode eigenvectors, 
each DOF $\eta$ is an oscillator or standing wave
with a different angular frequency ${\omega_\eta}$, in general, 
whose energy
can be expressed as 
\begin{align}
\energy_\eta\equiv\frac{1}{2}(\dot{Q}_\eta^2 + \omega_\eta^2 Q_\eta^2),
\label{eqn:energy}
\end{align}
where the {\em mode coordinate} ${Q_\eta}$ has the dimensions of ${\text{distance}\times\sqrt{\text{mass}}}$.
In the limit ${T\to 0}$ the behaviour of the system is described perfectly
by a Hamiltonian of the form
\begin{align}
\E(\bQ,\bP) 
& = U(\bQ^{\text{min}}) + \frac{1}{2}\sum_{\eta}\left[P_\eta^2 + \omega_\eta^2 Q_\eta^2\right],
\label{eqn:hamiltonian}
\end{align}
where ${U(\bQ^{\text{min}})}$ is a constant that is irrelevant to the dynamics, 
and ${P_\eta\equiv\dot{Q}_\eta}$ is the momentum conjugate to ${Q_\eta}$.

As illustrated in Fig.~\ref{fig:one}, the true path ${\partial\R_\eta}$
of mode ${\eta}$ in its phase space ${\phasespace_{\eta}}$ is
continuous. It is only the accessible information about the path that 
is quantized. As discussed in Sec.~\ref{section:boltzmann} and at
the beginning of Sec.~\ref{section:bose_einstein}, each point
${\Gamma_\eta\in\partial\R_\eta}$ is indistinguishable from 
all points within a neighbourhood of it, whose area is $\hq$.

Uncertainty manifests differently in the microstate probability distribution
depending on which set of coordinates is used to specify the microstate.
Having found that $p$ is a Maxwell-Boltzmann distribution
when standard position and momentum coordinates ${(P_{\eta},Q_{\eta})}$ are used, let
us now perform the canonical transformation
${(Q_{\eta},P_{\eta})\mapsto(\I_\eta,\vartheta_{\eta})}$, 
where ${(\I_\eta,\vartheta_{\eta})}$ are the action-angle variables~\cite{landau, arnold, lanczos}.
Then we will deduce the form of $p$ when the microstate is specfied
as ${\bGamma=(\bII,\bvartheta)\equiv\left(\I_{1},\I_{2},\cdots,\vartheta_{1},\vartheta_{2},\cdots\right)}$.

The action variable is defined as
\begin{align*}
\I_{\eta} \equiv \frac{1}{2\pi}\oint_{\partial\R_\eta}P_{\eta}\dd{Q_{\eta}} 
= \frac{1}{2\pi}\int\int_{\R_\eta} \dd{P_{\eta}}\wedge\dd{Q_{\eta}},
\end{align*}
where the first integral is performed around the closed continuous trajectory ${\partial\R_\eta}$ defined
by the equation ${\E_{\eta}(Q_{\eta},P_{\eta})=\energy_{\eta}}$ and depicted in Fig.~\ref{fig:one}(a).
The second expression, which involves an integral over the region $\R_\eta$ enclosed by the
elliptical path ${\partial\R_\eta}$, follows
from the generalized Stokes theorem.

It follows from the definition of ${\I_{\eta}}$ that ${2\pi\I_{\eta}}$ is the area enclosed by ${\partial\R_\eta}$.
From Eq.~\ref{eqn:energy}, it is easy to see that the semi-axes of ${\partial\R_\eta}$
are ${\sqrt{2\energy_{\eta}}/\omega_{\eta}}$ and ${\sqrt{2\energy_{\eta}}}$. Therefore, 
equating two expressions for the area enclosed gives
\begin{align*}
2\pi \I_{\eta}= \frac{2\pi }{\omega_{\eta}}\energy_{\eta} \implies \I_{\eta} = \frac{\energy_{\eta}}{\omega_{\eta}}.
\end{align*}
The reason for choosing ${\I_{\eta}}$ as one of our variables should now
be apparent: It allows us to express the new mode Hamiltonian as
\begin{align}
\E_{\eta}(\vartheta_{\eta},\I_{\eta}) = \E_{\eta}(\I_{\eta})=\I_{\eta}\omega_{\eta}.
\end{align}

If we now followed precisely the same procedure with the new coordinates as we used to derive the Maxwell-Boltzmann distribution
in Sec.~\ref{section:boltzmann}, we would reach Eq.~\ref{eqn:boltzmann}, with $\partition$ and ${\partition_{\eta}}$
expressed as sums over all ${\bGamma\equiv (\bII,\bvartheta)\in\G}$ and over all ${\Gamma_{\eta}\in\G_{\eta}}$, 
respectively. That is,
\begin{align*}
\partition = \sum_{\bGamma} e^{-\beta\E(\bGamma)}=\sum_{\bII}e^{-\beta\E(\bII)} = \prod_{\eta} \partition_{\eta},
\end{align*}
where 
\begin{align}
\partition_{\eta}\equiv \sum_{\I_{\eta}}e^{-\beta\E_{\eta}(\I_{\eta})} = \sum_{\I_{\eta}} e^{-\beta \I_{\eta}\omega_{\eta}},
\label{eqn:mode_partition2}
\end{align}
and the sum over ${\I_{\eta}}$ is a sum over a maximal set, ${\G_{\eta}\equiv \Delta\I_{\eta}\left(\integernonneg+\frac{1}{2}\right)}$, 
of mutually-distinguishable values of ${\I_{\eta}}$. The reason for the factor ${\frac{1}{2}}$
is that the lower bound, ${\frac{1}{2}\Delta\I_{\eta}}$, on the difference between mutually-distinguishable
values of ${\I_{\eta}}$ makes all points within the interval ${[0,\frac{1}{2}\Delta\I_{\eta})}$
indistinguishable from zero, and makes zero indistinguishable from all points in this interval. 
Therefore, the sum in Eq.~\ref{eqn:mode_partition2} can be viewed as ${1/\Delta\I_{\eta}}$ times
a Riemann sum over ${\realpos}$, which samples intervals of width ${\Delta\I_{\eta}}$ centered at 
${\frac{1}{2}\Delta\I_{\eta}}$,
${\frac{3}{2}\Delta\I_{\eta}}$,
${\frac{5}{2}\Delta\I_{\eta}}$, etc..

Now, since $\hq$ is a phase space area, 
the unavoidable uncertainty in the value of ${2\pi\I_\eta}$ must be ${\hq}$, 
and the unavoidable uncertainty in the value of ${\I_\eta}$ must be ${\hbarq\equiv \hq/(2\pi)}$.
Therefore the partition function can be expressed as
\begin{align*}
\partition_{\eta} & = \sum_{n_{\eta}\in\integernonneg} e^{-\beta \left(n_{\eta}+\frac{1}{2}\right) \hbarq \omega_{\eta}} 
\nonumber\\
& = \frac{e^{-\frac{1}{2}\beta\hbarq\omega_{\eta}}}{1-e^{-\beta\hbarq\omega_{\eta}}} 
= \frac{e^{\frac{1}{2}\beta\hbarq\omega_{\eta}}}{e^{\beta\hbarq\omega_{\eta}}-1},
\end{align*}
where the second line has been reached by using the fact that the right hand side of the first line is an infinite geometric series.
We can now express the free energy as
\begin{align*}
\free & = -\beta^{-1}\log \partition = -\beta^{-1}\sum_{\eta}\log \partition_{\eta} 
\nonumber\\
 & = \sum_{\eta}\left[\frac{1}{2}\hbarq\omega_{\eta} + k_B T\log\left(1-e^{-\beta\hbarq\omega_{\eta}}\right)\right].
\end{align*}
The term ${\frac{1}{2}\hbarq\omega_{\eta}}$ is commonly known as the {\em zero point energy} of mode ${\eta}$.

We can also calculate the expectation value, 
\begin{align}
\bar{n}_{\eta} \equiv \partition_{\eta}^{-1}\sum_{n_{\eta}\in\integernonneg} n_{\eta}e^{-\beta\left(n_{\eta}+\frac{1}{2}\right)\hbarq\omega_{\eta}},
\end{align}
of ${n_{\eta}}$ using Eq.~\ref{eqn:mode_partition2} as follows:
\begin{align*}
\pdv{\beta}\left(\sum_{n_{\eta}\in\integernonneg}e^{-\beta\left(n_{\eta}+\frac{1}{2}\right)\hbarq\omega_{\eta}}\right)
= \pdv{\beta}\left(\frac{e^{\frac{1}{2}\beta\hbarq\omega_{\eta}}}{e^{\beta\hbarq\omega_{\eta}}-1}\right).
\end{align*}
After taking the derivatives and simplifying, this can be expressed as
\begin{align*}
\bar{n}_{\eta} = \frac{1}{e^{\beta\hbarq\omega_{\eta}}-1}. 
\end{align*}
The integer ${n_{\eta}}$ is commonly referred to as the {\em occupation number}
of mode ${\eta}$ and ${\bar{n}_{\eta}}$ is its thermal average.

When the modes' amplitudes are large enough that they do interact, 
their energies and frequencies vary, their paths in their phase spaces are no longer elliptical,
and matters become more complicated. Nevertheless, 
simplifying assumptions are often justified, which allow a Bose-Einstein distribution
to be used as the basis for a statistical description of the system's
microstates and observables. 
For example, if the energy of mode ${\eta}$
is modulated by a mode ${\eta'}$ whose frequency is sufficiently
low (${\omega_{\eta'}\ll\omega_{\eta}}$), then ${\I_{\eta}}$ is
approximately {\em adiabatically invariant} under this modulation~\cite{landau, arnold,lanczos}, 
and the dominant effect of the interaction on mode ${\eta}$ is
to modulate its frequency.

As another example, when the interactions between modes are weak, 
the distribution of each mode's energy among frequencies is broadened
and shifted relative to its \templim limit.
Therefore, it still has a well defined mean frequency 
and mean energy, which allows the Bose-Einstein distribution 
to be used effectively in many cases.

\subsection{Generalizations to non-oscillatory systems}
I have now derived the Bose-Einstein distribution
for a classical system whose dynamics is a superposition of independent
harmonic oscillations.
My derivation made use of two properties of the system's Hamiltonian:
The first was that it could be expressed as a sum ${\E=\sum_\eta\E_\eta}$
of the Hamiltonians $\E_\eta$ of independent DOFs.
The second was that each $\E_\eta$ could be expressed as
an affine function of only one variable. For oscillations, 
this was achieved by transforming to action-angle variables, so that
each ${\E_\eta}$ took the form  ${\E_\eta(\I_\eta)=\I_\eta\omega_\eta}$. 
Since variations of ${\omega_\eta}$ are negligible 
when interactions are weak, ${\I_\eta}$ is effectively 
the only variable that appears in $\E_\eta$.

The Hamiltonians of many other kinds of physical systems,
composed of mutually-noninteracting DOFs, can be
transformed canonically into forms that allow the
Bose-Einstein distribution to be derived.
In principle it can be derived whenever there exists
a curve ${\gamma_\eta:\realpos\to\phasespace_\eta; t\mapsto\gamma_\eta(t)}$ 
in the phase space $\phasespace_\eta$ of each DOF such that energies of DOF $\eta$ are represented on ${\gamma_\eta}$ 
in the same proportions as they are represented in ${\phasespace_\eta}$.
To be more precise, energies should be represented on {\em maximal samplings}
of ${\gamma_\eta}$ in the same proportions as they are represented
on maximal samplings of ${\phasespace_\eta}$.

Once each $\E_\eta$ has been transformed canonically
into the form ${\E_\eta(X_\eta)=B_\eta+ C_\eta X_\eta}$, 
where ${X_\eta\in\realone}$ is a continuously-varying generalized coordinate or momentum, and $B_\eta$
and $C_\eta$ are constants, the full Hamiltonian becomes
\begin{align}
\E(\bX) &\equiv U(\bQ^{\text{min}}) + \sum_{\eta} B_\eta + \sum_{\eta} C_\eta X_\eta
\label{eqn:general_hamiltonian}
\end{align}
where ${\bX\equiv(X_1, X_2,\cdots)}$.
Let ${\frac{1}{2}\Delta X_\eta}$ denote the smallest difference between mutually-distinguishable
values of $X_\eta$; let ${\epsilon_\eta\equiv C_\eta\Delta X_\eta}$; and 
let
${D \equiv e^{-\beta\left(U(\bQ^{\text{min}})+\sum_\eta B_\eta\right)}}$.
Then the partition function can be expressed as
\begin{align*}
\partition 
&= D\prod_\eta \sum_{n_\eta\in\integerpos_0} e^{-\beta (n_\eta +\frac{1}{2})\epsilon_\eta},
\end{align*}
and it is straightforward to show that
${\bar{n}_\eta  = 1/\left(e^{\beta\epsilon_\eta}-1\right)}$.

One example of a system whose Hamiltonian can be transformed canonically into the
form of Eq.~\ref{eqn:general_hamiltonian} is an ideal gas. At any given point
in time, its Hamiltonian has the form, ${\E(\bP)\equiv\sum_\eta \E_\eta(P_\eta)=\frac{1}{2}\sum_\eta P_\eta^2}$, 
which is the Hamiltonian of a set of independent free particles. A free particle Hamiltonian can be transformed
canonically into a harmonic oscillator Hamiltonian~\cite{free_particle}; therefore, it can also
be transformed into action-angle coordinates.

As discussed in Sec.~\ref{section:lowT}, in the limit ${T\to 0}$ the Hamiltonian of \emph{every} classical
dynamical system either takes the same form as a set of weakly interacting harmonic
oscillators or as an ideal gas,  or as a combination of both. Therefore, all classical
dynamical systems that are subject to an uncertainty principle, ${\DQ\DP>\hq>0}$, obey
Bose-Einstein statistics in the \templim~limit as 
a consequence of the probability domain quantization discussed in Sec.~\ref{section:domain_quantization}.

As the derivation of the Bose-Einstein distribution presented in Sec.~\ref{section:oscillators} makes
clear, when an uncertainty principle applies, 
the Maxwell-Boltzmann and Bose-Einstein
distributions are perfectly compatible with one another 
in the \templim~limit. 
In that limit,
a classical system's energy distribution can be expressed either as
a Maxwell-Boltzmann distribution or as a Bose-Einstein distribution, depending
on which choice of coordinates and their conjugate momenta
are used to specify the microstate.

\section{Discussion}
I have shown that the Bose-Einstein distribution follows mathematically from
{\em probability domain quantization}, and that probability domain quantization 
is a consequence of the existence of a 
limit, $\hq$, on the precision with which a system's microstate can be determined experimentally.
Probability domain quantization does not imply that the microstates
of the underlying physical system are quantized. It implies a quantization of the information contained in
probability distributions that possess the quality of being testable empirically.

I have not justified my working assumption that a lower bound $\hq$ exists, but only demonstrated
that one of its consequences would be that all sufficiently-cold classical dynamical systems
are described by Bose-Einstein statistics. Therefore I have demonstrated that the
existence of such a lower bound would have many important implications.

One implication would be that there is no qualitative discrepancy between
the experimentally-observed spectrum of a blackbody
and what should be expected if light was a mechanical wave in a bounded medium.
As discussed in Sec.~\ref{section:uncertainty}, if light was such a wave, the boundedness
of the medium would mean that the smallest 
energy difference between two light waves of frequency ${\approx f}$ 
would be ${h_m f}$, for some constant $h_m$. In Sec.~\ref{section:bose_einstein} I showed
that the lower bound that an uncertainty principle would place on \emph{observable} 
energy differences would be ${\Delta\left(\I \omega\right)=\left(\Delta\I\right)\omega = \hq f}$.
Therefore, if ${h_m=\hq}$, the existence of an uncertainty principle in a classical universe 
could be explained by all observations being mediated by classical light waves.

Another implication of a lower bound $\hq$ would be that
there is no qualitative discrepancy between the experimentally-observed
temperature dependence of a crystal's heat capacity
and what should be expected of classical lattice waves. 

Another implication would be that classical oscillators and waves would have
\emph{zero point energies} that were simply an artefact
of small energies being empirically-indistinguishable from no energy.

Another implication would be that, when a cluster of massive particles was cold enough,
the classical expectation would be that almost all of its vibrational energy would be possessed
by its lowest-frequency normal mode. Therefore, below a certain temperature, all
but one of its degrees of freedom would be almost inactive and it would be a Bose-Einstein
condensate.

For simplicity I have assumed
that the limit on microstate measurement precision that leads to probability domain
quantization is a limit on \emph{certain} knowledge. 
In other words, I assumed that it is theoretically possible to know \emph{with certainty} that a DOF's microstate
is within a subset of its phase space if and only if the area of that subset
is greater than $\hq$.
If, instead, it is assumed that the result of the most precise microstate measurements possible
are probability density functions of the form
\begin{align*}
\rho(\SQ,\SP):\configspace\times\momspace\to\realpos;\; (Q,P)\mapsto \rho(Q,P;\SQ,\SP),
\end{align*}
where ${\SQ}$ and ${\SP}$ are the standard deviations along the coordinate axis ${\configspace}$
and the momentum axis ${\momspace}$, respectively, 
a more general form of uncertainty principle would be ${\SQ\SP>\hq}$.
This would be a limit on \emph{probabilistic} knowledge.
It may be possible to adapt the derivations presented in this work to uncertainty principles 
of this more general form.

\bibliography{quantumstatistics}
\end{document}